\documentclass[preprint,pre,showpacs,showkeys]{revtex4}%
\usepackage{amssymb}
\usepackage{amsmath}
\usepackage{amsfonts}
\usepackage{graphicx}%
\newtheorem{theorem}{Theorem}
\newtheorem{acknowledgement}[theorem]{Acknowledgement}

\begin{document}
\title[ ]{Mathematical structure\\derived from the $q$-multinomial coefficient in Tsallis statistics}
\author{Hiroki Suyari}
\email{suyari@ieee.org, suyari@faculty.chiba-u.jp}
\affiliation{Department of Information and Image Sciences, Faculty of Engineering, Chiba
University, 263-8522, Japan}

\begin{abstract}
We present the conclusive mathematical structure behind Tsallis statistics. We
obtain mainly the following five theoretical results: (i) the one-to-one
correspondence between the $q-$multinomial coefficient and Tsallis entropy,
(ii) symmetry behind Tsallis statistics, (iii) the numerical computations
revealing the existence of the central limit theorem in Tsallis statistics,
(iv) Pascal's triangle in Tsallis statistics and its properties, (v) the
self-similarity of the $q-$product $\otimes_{q}$ leading to successful
applications in Tsallis statistics. In particular, the third result (iii)
provides us with a mathematical representation of a convincible answer to the
physical problem: \textquotedblleft Why so many power-law behaviors exist in
nature universally ?\textquotedblright

\end{abstract}
\eid{ }
\date{\today }
\startpage{1}
\endpage{1}
\maketitle

\section{Introduction}

Tsallis entropy introduced in 1988 \cite{Ts88}\cite{CuTs91}:%
\begin{equation}
S_{q}\left(  p_{1},\cdots,p_{n}\right)  :=\frac{1-\sum\limits_{i=1}^{n}%
p_{i}^{q}}{q-1}\label{Tsallis entropy}%
\end{equation}
has been considered to obtain and provide new possibilities to construct the
generalized statistical physics recovering not only the traditional
Boltzmann-Gibbs statistical mechanics but also the so-called Tsallis
statistics to describe power-law behaviors systematically. Until now, the
maximum entropy principle (MEP for short) has been mainly applied to
mathematical foundations in Tsallis statistics along the same lines of Jaynes'
ideas \cite{AO01}\cite{Ja83}. The MEP for Tsallis statistics yields
equilibrium states exhibiting power-law behaviors, which provide us many
satisfactory descriptions of complex systems such as self-gravitating systems
\cite{PP93}\cite{Ta02}, pure electron plasma \cite{B96}\cite{AT97}, full
developed turbulence \cite{AA00}\cite{Be00}, high-energy collisions
\cite{WW00}\cite{WR00} and so on. Through the history of sciences, we have
learned an important lesson that there always exists a beautiful mathematical
structure at the birth of a new physics. This lesson stimulates us to finding
it behind Tsallis statistics. On the way to the goal, we obtain some
theoretical results coming from the mathematical structure in Tsallis
statistics. The key concept leading to our results is \textquotedblleft the
$q-$\textit{product\textquotedblright} uniquely determined by Tsallis entropy,
which is first introduced by Borges in \cite{Bo03}.

The $q-$\textit{product} determined by Tsallis entropy has been successfully
applied to the derivations of \textquotedblleft\textit{the law of error
\cite{ST0312}}\textquotedblright\ and \textquotedblleft\textit{the }%
$q-$\textit{Stirling's formula\cite{ST04a}}\textquotedblright\ in Tsallis
statistics. The $q-$product is defined as follows:%
\begin{equation}
x\otimes_{q}y:=\left\{
\begin{array}
[c]{ll}%
\left[  x^{1-q}+y^{1-q}-1\right]  ^{\frac{1}{1-q}}, & \text{if }%
x>0,\,y>0,\,x^{1-q}+y^{1-q}-1>0,\\
0, & \text{otherwise.}%
\end{array}
\right. \label{def of q-product}%
\end{equation}
The definition of the $q-$product originates from the requirement of the
following satisfactions:%
\begin{align}
\ln_{q}\left(  x\otimes_{q}y\right)   &  =\ln_{q}x+\ln_{q}%
y,\label{property of ln_q}\\
\exp_{q}\left(  x\right)  \otimes_{q}\exp_{q}\left(  y\right)   &  =\exp
_{q}\left(  x+y\right)  ,\label{property of exp_q}%
\end{align}
\ where $\ln_{q}x$ and $\exp_{q}\left(  x\right)  $ are the $q-$%
\textit{logarithm function}:%
\begin{equation}
\ln_{q}x:=\frac{x^{1-q}-1}{1-q}\quad\left(  x>0,q\in\mathbb{R}^{+}\right)
\end{equation}
and its inverse function, the $q-$\textit{exponential function}:
\begin{equation}
\exp_{q}\left(  x\right)  :=\left\{
\begin{array}
[c]{ll}%
\left[  1+\left(  1-q\right)  x\right]  ^{\frac{1}{1-q}} & \text{if }1+\left(
1-q\right)  x>0,\\
0 & \text{otherwise}%
\end{array}
\right.  \quad\left(  x\in\mathbb{R},q\in\mathbb{R}^{+}\right)
.\label{def of expq}%
\end{equation}
\textit{\ }These functions, $\ln_{q}x$ and $\exp_{q}\left(  x\right)  $, are
originally determined by Tsallis entropy and its maximization \cite{AO01}%
\cite{Ts94}. As shown in our previous paper \cite{ST04a}\textit{, }$\exp
_{q}\left(  x\right)  $ is rewritten by means of the $q-$product.%
\begin{equation}
\exp_{q}\left(  x\right)  =\underset{n\rightarrow\infty}{\lim}\left(
1+\frac{x}{n}\right)  ^{\otimes_{q}^{n}}\label{re-def of expq}%
\end{equation}
This representation (\ref{re-def of expq}) is a natural generalization of the
famous definition of the usual exponential function: $\exp\left(  x\right)
=\underset{n\rightarrow\infty}{\lim}\left(  1+\frac{x}{n}\right)  ^{n}$. This
fundamental property (\ref{re-def of expq}) reveals the conclusive validity of
the $q-$product in Tsallis statistics. See the appendix of \cite{ST04a} for
the proof.

Besides the fundamental property (\ref{re-def of expq}), until now we have
presented two successful applications of the $q-$product in Tsallis
statistics: \textquotedblleft\textit{the law of error\cite{ST0312}%
}\textquotedblright\ and \textquotedblleft\textit{the }$q-$\textit{Stirling's
formula\cite{ST04a}}\textquotedblright.

In the former application \cite{ST0312}, the $q-$product is applied to the
likelihood function $L_{q}\left(  \theta\right)  $ in maximum likelihood
principle (MLP for short) instead of the usual product.
\begin{equation}
L_{q}\left(  \theta\right)  =L_{q}\left(  x_{1},x_{2},\cdots,x_{n}%
;\theta\right)  :=f\left(  x_{1}-\theta\right)  \otimes_{q}f\left(
x_{2}-\theta\right)  \otimes_{q}\cdots\otimes_{q}f\left(  x_{n}-\theta\right)
\end{equation}
where $x_{1},x_{2},\cdots,x_{n}\in\mathbb{R}$ are results of $n$ measurements
for some observation. The requirement for taking the maximal value of
$L_{q}\left(  \theta\right)  $ at%
\begin{equation}
\theta=\theta^{\ast}:=\frac{x_{1}+x_{2}+\cdots+x_{n}}{n}%
\end{equation}
uniquely determines a \textit{Tsallis distribution}:%
\begin{equation}
f\left(  e\right)  =\frac{\exp_{q}\left(  -\beta_{q}e^{2}\right)  }{\int
\exp_{q}\left(  -\beta_{q}e^{2}\right)  de}\quad\left(  \beta_{q}>0\right)
\end{equation}
as a nonextensive generalization of a Gaussian distribution.

In the latter application \cite{ST04a}, using the $q-$product, we obtain the
$q-$Stirling's formula. For the $q-$factorial $n!_{q}\,$\ for $n\in\mathbb{N}$
and $q>0$ defined by%
\begin{equation}
n!_{q}:=1\otimes_{q}\cdots\otimes_{q}n,
\end{equation}
the $q-$Stirling's formula $\left(  q\neq1\right)  $ is%
\begin{equation}
\ln_{q}\left(  n!_{q}\right)  \simeq\left\{
\begin{array}
[c]{ll}%
\left(  \frac{n}{2-q}+\frac{1}{2}\right)  \frac{n^{1-q}-1}{1-q}+\left(
-\frac{n}{2-q}\right)  +\left(  \frac{1}{2-q}-\delta_{q}\right)  &
\text{if}\quad q>0\quad\text{and}\quad q\neq1,2\\
n-\frac{1}{2n}-\ln n-\frac{1}{2}-\delta_{2} & \text{if}\quad q=2
\end{array}
\right.
\end{equation}
where $\delta_{q}$ is a $q-$dependent parameter which does \textit{not} depend
on $n$. Slightly rough expression of the $q-$Stirling's formula $\left(
q\neq1\right)  $ is%
\begin{equation}
\ln_{q}\left(  n!_{q}\right)  \simeq\left\{
\begin{array}
[c]{ll}%
\frac{n}{2-q}\left(  \ln_{q}n-1\right)  & \text{if}\quad q>0\quad
\text{and}\quad q\neq1,2\\
n-\ln n & \text{if}\quad q=2
\end{array}
\right.  .\label{rough q-Stirling}%
\end{equation}
On the other hand, a more strict expression of the $q-$Stirling's formula
$\left(  q\neq1\right)  $ are%
\begin{align}
&  \frac{n}{2-q}\ln_{q}n-\frac{n}{2-q}+\frac{1}{2}\ln_{q}n+\frac{1}{2-q}%
-\frac{1}{8}<\ln_{q}\left(  n!_{q}\right) \nonumber\\
\!\! &  <\frac{n}{2-q}\ln_{q}n-\frac{n}{2-q}+\frac{1}{2}\ln_{q}n+\frac{1}%
{2-q}\text{\quad}\left(  \text{if}\quad q>0\quad\text{and}\quad q\neq
1,2\right)
\end{align}
and%
\begin{equation}
n-\ln n-\frac{1}{2n}-\frac{5}{8}<\ln_{q}\left(  n!_{q}\right)  <n-\ln
n-\frac{1}{2n}-\frac{1}{2}\text{\quad}\left(  \text{if}\quad q=2\right)  .
\end{equation}
These $q-$Stirling's formulas recover the famous Stirling's formulas when
$q\rightarrow1.$

\bigskip Along the lines of these successful applications of the $q-$product,
we introduce \textit{the }$q-$\textit{multinomial coefficient} using the
$q-$product in section II. The present $q-$multinomial coefficient and the
Tsallis entropy has a surprising relationship between them, which indicates a
symmetry behind Tsallis statistics. Moreover, the numerical computations of
the $q-$binomial coefficients are shown in section III. When $n$ goes
infinity, a set of \bigskip the normalized $q-$binomial coefficients converge
to its corresponding Tsallis distribution with the same $q$. These numerical
results indicate the existence of the central limit theorem in Tsallis
statistics as a nonextensive generalization of that of the usual probability
theory. In section IV, we reveal the reason why the $q$-product can be
successfully applied to Tsallis statistics. The self-similarity of the
$q-$product plays essential roles for these successful applications in Tsallis
statistics. In the appendix, Pascal's triangle in Tsallis statistics and its
properties are presented using the present $q-$binomial coefficient.

\section{$q-$multinomial coefficient in Tsallis statistics and its one-to-one
correspondence to Tsallis entropy}

\bigskip We define the $q-$binomial coefficient $\left[
\begin{array}
[c]{c}%
n\\
k
\end{array}
\right]  _{q}$ by%
\begin{equation}
\left[
\begin{array}
[c]{c}%
n\\
k
\end{array}
\right]  _{q}:=\left(  n!_{q}\right)  \oslash_{q}\left[  \left(
k!_{q}\right)  \otimes_{q}\left(  \left(  n-k\right)  !_{q}\right)  \right]
\quad\quad\left(  n,k\left(  \leq n\right)  \in\mathbb{N}\right)
\label{def of q-binomial coefficient}%
\end{equation}
where $\oslash_{q}$ is the inverse operation to $\otimes_{q}$, which is
defined by
\begin{equation}
x\oslash_{q}y:=\left\{
\begin{array}
[c]{lll}%
\left[  x^{1-q}-y^{1-q}+1\right]  ^{\frac{1}{1-q}}, &  & \text{if
}x>0,\,y>0,\,x^{1-q}-y^{1-q}+1>0,\\
0, &  & \text{otherwise}%
\end{array}
\right.  .
\end{equation}
$\oslash_{q}$ is also introduced by the following satisfactions as similarly
as $\otimes_{q}$ \cite{Bo03}.
\begin{align}
\ln_{q}x\oslash_{q}y &  =\ln_{q}x-\ln_{q}y,\\
\exp_{q}\left(  x\right)  \oslash_{q}\exp_{q}\left(  y\right)   &  =\exp
_{q}\left(  x-y\right)  .
\end{align}
Applying the definitions of $\otimes_{q}$, $\oslash_{q}$ and $n!_{q}$ given
by
\begin{equation}
n!_{q}=1\otimes_{q}\cdots\otimes_{q}n=\left[  \sum_{k=1}^{n}k^{1-q}-\left(
n-1\right)  \right]  ^{\frac{1}{1-q}}\label{q-kaijyo}%
\end{equation}
to (\ref{def of q-binomial coefficient}), the $q-$binomial coefficient is
explicitly written as
\begin{equation}
\left[
\begin{array}
[c]{c}%
n\\
k
\end{array}
\right]  _{q}=\left[  \sum_{\ell=1}^{n}\ell^{1-q}-\sum_{i=1}^{k}i^{1-q}%
-\sum_{j=1}^{n-k}j^{1-q}+1\right]  ^{\frac{1}{1-q}}.\label{expr of q-binomial}%
\end{equation}
In general, when $q<1$, $\sum_{\ell=1}^{n}\ell^{1-q}-\sum_{i=1}^{k}%
i^{1-q}-\sum_{j=1}^{n-k}j^{1-q}+1>0$. Thus, we can plot the probability
distribution determined by the normalized $q-$binomial coefficients in case
$0<q<1$. (See Fig.1). From the definition (\ref{def of q-binomial coefficient}%
), it is clear that%
\begin{equation}
\underset{q\rightarrow1}{\lim}\left[
\begin{array}
[c]{c}%
n\\
k
\end{array}
\right]  _{q}=\left[
\begin{array}
[c]{c}%
n\\
k
\end{array}
\right]  =\frac{n!}{k!\left(  n-k\right)  !}.
\end{equation}
On the other hand, when $\sum_{\ell=1}^{n}\ell^{1-q}-\sum_{i=1}^{k}%
i^{1-q}-\sum_{j=1}^{n-k}j^{1-q}+1<0,$ $\left[
\begin{array}
[c]{c}%
n\\
k
\end{array}
\right]  _{q}$ takes complex numbers in general, which divides the
formulations and discussions of the $q-$binomial coefficient $\left[
\begin{array}
[c]{c}%
n\\
k
\end{array}
\right]  _{q}$ into two cases: it takes a real number or a complex number. In
order to avoid such separate formulations and discussions,\ we consider the
$q$-logarithm of the $q-$binomial coefficient:
\begin{equation}
\ln_{q}\left[
\begin{array}
[c]{c}%
n\\
k
\end{array}
\right]  _{q}=\ln_{q}\left(  n!_{q}\right)  -\ln_{q}\left(  k!_{q}\right)
-\ln_{q}\left(  \left(  n-k\right)  !_{q}\right)
.\label{epr of lnq of q-binomial}%
\end{equation}
For simplicity, we consider the only case $q\neq1$ and $q>0$ throughout the
paper.%
\begin{figure}
[ptbh]
\begin{center}
\includegraphics[
height=3.7429in,
width=4.9562in
]%
{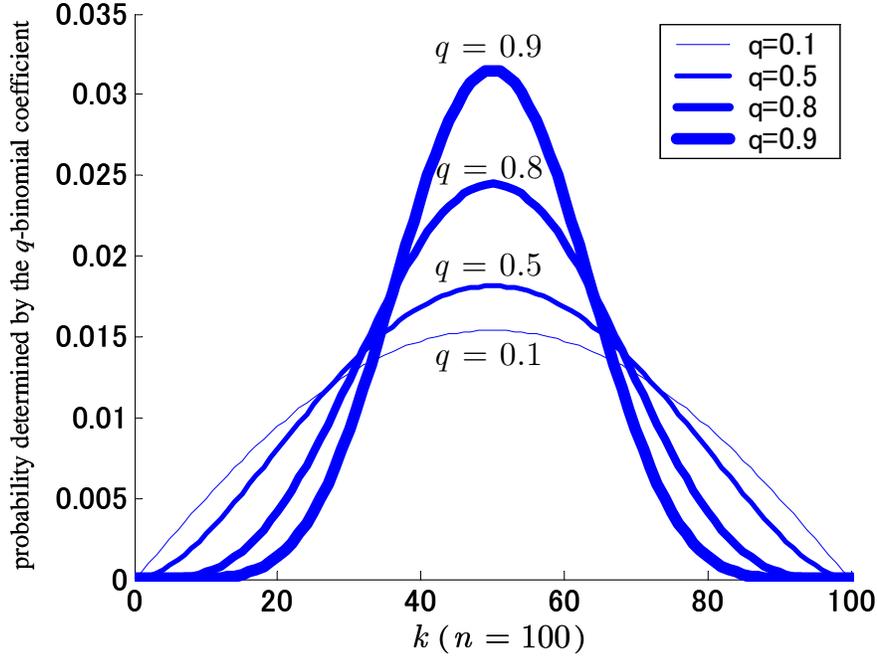}%
\caption{probability distribution determined by the normalized $q-$binomial
coefficients}
\label{fig: qbinomial_qnosuii}%
\end{center}
\end{figure}

\bigskip The above definition (\ref{def of q-binomial coefficient}) is
artificial because it is defined from the analogy with the usual binomial
coefficient $\left[
\begin{array}
[c]{c}%
n\\
k
\end{array}
\right]  $. However, when $n$ goes infinity, the $q-$binomial coefficient
(\ref{def of q-binomial coefficient}) has a surprising relation to Tsallis
entropy as follows:%
\begin{equation}
\ln_{q}\left[
\begin{array}
[c]{c}%
n\\
k
\end{array}
\right]  _{q}\simeq\left\{
\begin{array}
[c]{ll}%
\frac{n^{2-q}}{2-q}\cdot S_{2-q}\left(  \frac{k}{n},\frac{n-k}{n}\right)  &
\text{if}\quad q>0,\,\,q\neq2\\
-S_{1}\left(  n\right)  +S_{1}\left(  k\right)  +S_{1}\left(  n-k\right)  &
\text{if}\quad q=2
\end{array}
\right.  ,\label{important0-1}%
\end{equation}
where $S_{q}$ is Tsallis entropy (\ref{Tsallis entropy}) and $S_{1}\left(
n\right)  $ is Boltzmann entropy:%
\begin{equation}
S_{1}\left(  n\right)  :=\ln n.
\end{equation}

Applying the rough expression of the $q-$Stirling's formula
(\ref{rough q-Stirling}), the above relations (\ref{important0-1}) are easily
proved as follows:%
\begin{align}
\text{\quad if}\quad q>0,\,\,q\neq2, & \nonumber\\
\ln_{q}\left[
\begin{array}
[c]{c}%
n\\
k
\end{array}
\right]  _{q} &  \simeq\frac{n}{2-q}\left(  \ln_{q}n-1\right)  -\frac{k}%
{2-q}\left(  \ln_{q}k-1\right)  -\frac{n-k}{2-q}\left(  \ln_{q}\left(
n-k\right)  -1\right) \nonumber\\
&  =\frac{n}{2-q}\ln_{q}n-\frac{k}{2-q}\ln_{q}k-\frac{n-k}{2-q}\ln_{q}\left(
n-k\right) \nonumber\\
&  =\frac{n^{2-q}-k^{2-q}-\left(  n-k\right)  ^{2-q}}{\left(  2-q\right)
\left(  1-q\right)  }\nonumber\\
&  =\frac{n^{2-q}}{2-q}\cdot\frac{1-\left(  \frac{k}{n}\right)  ^{2-q}-\left(
\frac{n-k}{n}\right)  ^{2-q}}{\left(  2-q\right)  -1}\nonumber\\
&  =\frac{n^{2-q}}{2-q}\cdot S_{2-q}\left(  \frac{k}{n},\frac{n-k}{n}\right)
,\\
\text{\quad if}\quad\,q=2,\text{ \ \ \ \ \ \ \ \ \ \ } & \nonumber\\
\ln_{q}\left[
\begin{array}
[c]{c}%
n\\
k
\end{array}
\right]  _{q} &  \simeq\left(  n-\ln n\right)  -\left(  k-\ln k\right)
-\left(  \left(  n-k\right)  -\ln\left(  n-k\right)  \right) \nonumber\\
&  =-\ln n+\ln k+\ln\left(  n-k\right) \nonumber\\
&  =-S_{1}\left(  n\right)  +S_{1}\left(  k\right)  +S_{1}\left(  n-k\right)
.
\end{align}

The above correspondence (\ref{important0-1}) between the $q-$binomial
coefficient (\ref{def of q-binomial coefficient}) and Tsallis entropy
convinces us of the fact the $q-$binomial coefficient
(\ref{def of q-binomial coefficient}) is well-defined in Tsallis statistics.
Therefore, we can construct Pascal's triangle in Tsallis statistics. See the
appendix for the detail.

The above relation (\ref{important0-1}) is easily generalized to the case of
the $q-$multinomial coefficient. The $q-$multinomial coefficient in Tsallis
statistics is defined in a similar way as that of the the $q-$binomial
coefficient (\ref{def of q-binomial coefficient}).%
\begin{equation}
\left[
\begin{array}
[c]{ccc}
& n & \\
n_{1} & \cdots & n_{k}%
\end{array}
\right]  _{q}:=\left(  n!_{q}\right)  \oslash_{q}\left[  \left(  n_{1}%
!_{q}\right)  \otimes_{q}\cdots\otimes_{q}\left(  n_{k}!_{q}\right)  \right]
\label{def of q-multinomial coefficient}%
\end{equation}
where
\begin{equation}
n=\sum_{i=1}^{k}n_{i},\quad n_{i}\in\mathbb{N\,}\left(  i=1,\cdots,k\right)  .
\end{equation}
Applying the definitions of $\otimes_{q}$ and $\oslash_{q}$ to
(\ref{def of q-multinomial coefficient}), the $q-$multinomial coefficient is
explicitly written as
\begin{equation}
\left[
\begin{array}
[c]{ccc}
& n & \\
n_{1} & \cdots & n_{k}%
\end{array}
\right]  _{q}=\left[  \sum_{\ell=1}^{n}\ell^{1-q}-\sum_{i_{1}=1}^{n_{1}}%
i_{1}^{1-q}\cdots-\sum_{i_{k}=1}^{n_{k}}i_{k}^{1-q}+1\right]  ^{\frac{1}{1-q}%
}.\label{expr of q-multinomial}%
\end{equation}
Along the same reason as stated above in case of the $q-$binomial coefficient,
we consider the $q$-logarithm of the $q-$multinomial coefficient given by
\begin{equation}
\ln_{q}\left[
\begin{array}
[c]{ccc}
& n & \\
n_{1} & \cdots & n_{k}%
\end{array}
\right]  _{q}=\ln_{q}\left(  n!_{q}\right)  -\ln_{q}\left(  n_{1}!_{q}\right)
\cdots-\ln_{q}\left(  n_{k}!_{q}\right)  .
\end{equation}

From the definition (\ref{def of q-multinomial coefficient}), it is clear that%
\begin{equation}
\underset{q\rightarrow1}{\lim}\left[
\begin{array}
[c]{ccc}
& n & \\
n_{1} & \cdots & n_{k}%
\end{array}
\right]  _{q}=\left[
\begin{array}
[c]{ccc}
& n & \\
n_{1} & \cdots & n_{k}%
\end{array}
\right]  =\frac{n!}{n_{1}!\cdots n_{k}!}.
\end{equation}
When $n$ goes infinity, the $q-$multinomial coefficient
(\ref{def of q-multinomial coefficient}) has the similar relation to Tsallis
entropy as (\ref{important0-1}):%
\begin{equation}
\ln_{q}\left[
\begin{array}
[c]{ccc}
& n & \\
n_{1} & \cdots & n_{k}%
\end{array}
\right]  _{q}\simeq\left\{
\begin{array}
[c]{ll}%
\frac{n^{2-q}}{2-q}\cdot S_{2-q}\left(  \frac{n_{1}}{n},\cdots,\frac{n_{k}}%
{n}\right)  & \text{if}\quad q>0,\,\,q\neq2\\
-S_{1}\left(  n\right)  +\sum\limits_{i=1}^{k}S_{1}\left(  n_{i}\right)  &
\text{if}\quad q=2
\end{array}
\right.  .\label{important0-2}%
\end{equation}
This is a natural generalization of (\ref{important0-1}). In the same way as
the case of the $q-$binomial coefficient, the above relation
(\ref{important0-2}) is proved as follows:%
\begin{align}
\text{\quad if}\quad q>0,\,\,q\neq2,\quad\quad & \nonumber\\
\ln_{q}\left[
\begin{array}
[c]{ccc}
& n & \\
n_{1} & \cdots & n_{k}%
\end{array}
\right]  _{q} &  \simeq\frac{n}{2-q}\left(  \ln_{q}n-1\right)  -\frac{n_{1}%
}{2-q}\left(  \ln_{q}n_{1}-1\right)  -\cdots-\frac{n_{k}}{2-q}\left(  \ln
_{q}n_{k}-1\right) \nonumber\\
&  =\frac{n}{2-q}\ln_{q}n-\frac{n_{1}}{2-q}\ln_{q}n_{1}-\cdots-\frac{n_{k}%
}{2-q}\ln_{q}n_{k}\nonumber\\
&  =\frac{n^{2-q}-n_{1}^{2-q}-\cdots-n_{k}^{2-q}}{\left(  2-q\right)  \left(
1-q\right)  }\nonumber\\
&  =\frac{n^{2-q}}{2-q}\cdot\frac{1-\left(  \frac{n_{1}}{n}\right)
^{2-q}-\cdots-\left(  \frac{n_{k}}{n}\right)  ^{2-q}}{\left(  2-q\right)
-1}\nonumber\\
&  =\frac{n^{2-q}}{2-q}\cdot S_{2-q}\left(  \frac{n_{1}}{n},\cdots,\frac
{n_{k}}{n}\right)  ,\nonumber\\
\text{if}\quad q=2,\quad\quad\quad\quad\quad\quad & \\
\ln_{q}\left[
\begin{array}
[c]{ccc}
& n & \\
n_{1} & \cdots & n_{k}%
\end{array}
\right]  _{q} &  \simeq\left(  n-\ln n\right)  -\left(  n_{1}-\ln
n_{1}\right)  -\cdots-\left(  n_{k}-\ln n_{k}\right) \nonumber\\
&  =-\ln n+\ln n_{1}+\cdots+\ln n_{k}\nonumber\\
&  =-S_{1}\left(  n\right)  +\sum\limits_{i=1}^{k}S_{1}\left(  n_{i}\right)  .
\end{align}

When $q\rightarrow1$, (\ref{important0-2}) recovers the well known result
\cite{CT91}:%
\begin{equation}
\ln\left[
\begin{array}
[c]{ccc}
& n & \\
n_{1} & \cdots & n_{k}%
\end{array}
\right]  \simeq nS_{1}\left(  \frac{n_{1}}{n},\cdots,\frac{n_{k}}{n}\right)
\end{equation}
where $S_{1}$ is Shannon entropy.

The present relation (\ref{important0-2}) tells us some significant messages
about Tsallis statistics. In particular, we take the following three of them.

\begin{enumerate}
\item There always exists a \textit{one-to-one correspondence} between Tsallis
entropy and the $q-$multinomial coefficient. In particular, from
(\ref{important0-2}) we obtain the following equivalence which makes sense in
statistical physics.
\begin{align}
&  ``\text{Maximization of }S_{2-q}\left(  \frac{n_{1}}{n},\cdots,\frac{n_{k}%
}{n}\right)  \text{ is equivalent to }\nonumber\\
&  \text{that of the }q\text{-multinomial coefficient }\left[
\begin{array}
[c]{ccc}
& n & \\
n_{1} & \cdots & n_{k}%
\end{array}
\right]  _{q}\text{ \ when }q<2\text{ and }n\text{ is large.
\textquotedblright}\label{l=2}%
\end{align}

\item When $q\neq2,$ the relation (\ref{important0-2}) is rewritten by%
\begin{equation}
\ln_{q}\left[
\begin{array}
[c]{ccc}
& n & \\
n_{1} & \cdots & n_{k}%
\end{array}
\right]  _{q}\simeq\frac{n^{q\prime}}{q\prime}\cdot S_{q\prime}\left(
\frac{n_{1}}{n},\cdots,\frac{n_{k}}{n}\right) \label{range of q-1}%
\end{equation}
where%
\begin{equation}
q+q\prime=2,\quad q>0.\label{range of q-2}%
\end{equation}
\quad In general, when Tsallis entropy $S_{q\prime}$ is applied, we consider
the case $q\prime>0$ only, because Tsallis entropy when $q\prime\leq0$ loses
its concavity which plays important roles in nonextensive systems. Thus,%
\begin{equation}
``q\prime=2-q>0,\quad q>0\text{ \textquotedblright\quad}\Leftrightarrow
\text{\quad}``0<q<2,\quad0<q\prime=2-q<2\ \text{ \textquotedblright}.
\end{equation}
This is in good agreement with the fact that the parameter $q\prime$ of
Tsallis entropy $S_{q\prime}$ takes a value in $\left(  0,2\right)  $ in most
of the successful physical models using Tsallis entropy. More precisely, in
case of $q\in\left(  0,1\right]  $ $\left(  \text{i.e., }q\prime\in\left[
1,2\right)  \right)  ,$ the $q$-multinomial coefficient $\left[
\begin{array}
[c]{ccc}
& n & \\
n_{1} & \cdots & n_{k}%
\end{array}
\right]  _{q}$ \textit{takes real numbers} because $\sum_{\ell=1}^{n}%
\ell^{1-q}-\sum_{i=1}^{n_{1}}i^{1-q}\cdots-\sum_{j=1}^{n_{k}}j^{1-q}+1>0$ in
(\ref{expr of q-multinomial}). On the other hand, in case of $q\in\left(
1,2\right)  $ $\left(  \text{i.e., }q\prime\in\left(  0,1\right)  \right)  ,$
the $q$-logarithm of the $q$-multinomial coefficients \textit{take real
numbers}, but the $q$-multinomial coefficient \textit{does not take real
numbers} in general because $\sum_{\ell=1}^{n}\ell^{1-q}-\sum_{i=1}^{n_{1}%
}i^{1-q}\cdots-\sum_{j=1}^{n_{k}}j^{1-q}+1<0$ in (\ref{expr of q-multinomial}%
). Thus, we summarize these considerations as follows.
\begin{align}
&  q\in\left(  0,1\right]  \quad\left(  \text{i.e., }q\prime\in\left[
1,2\right)  \right) \nonumber\\
&  \Rightarrow\text{the }q\text{-multinomial coefficients }\left[
\begin{array}
[c]{ccc}
& n & \\
n_{1} & \cdots & n_{k}%
\end{array}
\right]  _{q}\text{ uniquely corresponding to \ }\nonumber\\
&  \text{Tsallis entropy }S_{q\prime}=S_{2-q}\text{ \textit{take real numbers}
as their values,}\label{range1}\\
& \nonumber\\
&  q\in\left(  1,2\right)  \quad\left(  \text{i.e., }q\prime\in\left(
0,1\right)  \right) \nonumber\\
&  \Rightarrow\text{the }q\text{-multinomial coefficients }\left[
\begin{array}
[c]{ccc}
& n & \\
n_{1} & \cdots & n_{k}%
\end{array}
\right]  _{q}\text{ uniquely corresponding to }\nonumber\\
\  &  \text{Tsallis entropy }S_{q\prime}=S_{2-q}\text{ \textit{does not take
real numbers} as their values in general,}\nonumber\\
&  \text{but }\ln_{q}\left[
\begin{array}
[c]{ccc}
& n & \\
n_{1} & \cdots & n_{k}%
\end{array}
\right]  _{q}\text{ \textit{take real numbers}.}\label{range2}%
\end{align}
Therefore, in the next section, we present the numerical computations of the
$q$-binomial coefficients in case of $q\in\left(  0,1\right]  .$

\item The relation (\ref{important0-2}) reveals a surprising geometrical
structure: (\ref{important0-2}) is equivalent to%
\begin{equation}
\ln_{1-\left(  1-q\right)  }\left[
\begin{array}
[c]{ccc}
& n & \\
n_{1} & \cdots & n_{k}%
\end{array}
\right]  _{1-\left(  1-q\right)  }\simeq\frac{n^{1+\left(  1-q\right)  }%
}{1+\left(  1-q\right)  }\cdot S_{1+\left(  1-q\right)  }\left(  \frac{n_{1}%
}{n},\cdots,\frac{n_{k}}{n}\right)  \qquad\left(  \,q>0,\,\,q\neq2\right)
.\label{symmetry}%
\end{equation}
This expression represents that behind Tsallis statistics there exists a
\textit{symmetry} with a factor $1-q$ around $q=1$. Substitution of some
concrete values of $q$ such as $q=0.2,1.7$ into (\ref{important0-2}) helps us
understand the symmetry mentioned above.%
\begin{align}
&  \ln_{0.2}\left[
\begin{array}
[c]{ccc}
& n & \\
n_{1} & \cdots & n_{k}%
\end{array}
\right]  _{0.2}\simeq\frac{n^{1.8}}{1.8}\cdot S_{1.8}\left(  \frac{n_{1}}%
{n},\cdots,\frac{n_{k}}{n}\right) \nonumber\\
\text{\quad} &  \Leftrightarrow\text{\quad}\ln_{1-0.8}\left[
\begin{array}
[c]{ccc}
& n & \\
n_{1} & \cdots & n_{k}%
\end{array}
\right]  _{1-0.8}\simeq\frac{n^{1+0.\,8}}{1+0.8}\cdot S_{1+0.8}\left(
\frac{n_{1}}{n},\cdots,\frac{n_{k}}{n}\right)  \text{,}\nonumber\\
& \\
&  \ln_{1.7}\left[
\begin{array}
[c]{ccc}
& n & \\
n_{1} & \cdots & n_{k}%
\end{array}
\right]  _{1.7}\simeq\frac{n^{0.3}}{0.3}\cdot S_{0.3}\left(  \frac{n_{1}}%
{n},\cdots,\frac{n_{k}}{n}\right) \nonumber\\
\text{\quad} &  \Leftrightarrow\text{\quad}\ln_{1+0.7}\left[
\begin{array}
[c]{ccc}
& n & \\
n_{1} & \cdots & n_{k}%
\end{array}
\right]  _{1+0.7}\simeq\frac{n^{1-0.7}}{1-0.7}\cdot S_{1-0.7}\left(
\frac{n_{1}}{n},\cdots,\frac{n_{k}}{n}\right)  .\nonumber\\
&
\end{align}
Such a symmetry in Tsallis statistics has been discussed in the framework of
the relation to $q$-analysis \cite{Ab97}\cite{Jo98}. However, the present
symmetry is \textit{directly} derived from the definition of Tsallis entropy
as shown in this section.
\end{enumerate}

\section{Numerical computations of a set of the normalized $q-$binomial
coefficients and its convergence}

It is well known that any binomial distribution converge a Gaussian
distribution when $n$ goes infinity. This is a typical example of the central
limit theorem in the usual probability theory. By analogy with this famous
result, each set of normalized $q-$binomial coefficients is expected to
converge each Tsallis distribution with the same $q$ when $n$ goes infinity.
As shown in this section, the present numerical results come up to our expectations.

In Fig.2, each set of bars and solid line represent each set of normalized
$q-$binomial coefficients and Tsallis distribution with normalized $q$-mean $0
$ and normalized $q$-variance $1$ for each $n$ when $q=0.1$, respectively.
Each of the three graphs on the first row of Fig.2 represents two kinds of
probability distributions stated above, and the three graphs on the second row
of Fig.2 represent the corresponding cumulative probability distributions,
respectively. From Fig.2, we find the convergence of a set of normalized
$q-$binomial coefficients to a Tsallis distribution when $n$ goes infinity.
Fig.3 and Fig.4 represent the similar convergences in case of $q=0.5$ and
$q=0.9$ as the case of $q=0.1$, respectively.

Note that we \textit{never} use any curve-fitting in these numerical
computations. Every bar and solid line is computed and plotted independently
each other.%
\begin{figure}
[ptbh]
\begin{center}
\includegraphics[
height=5.2235in,
width=6.9548in
]%
{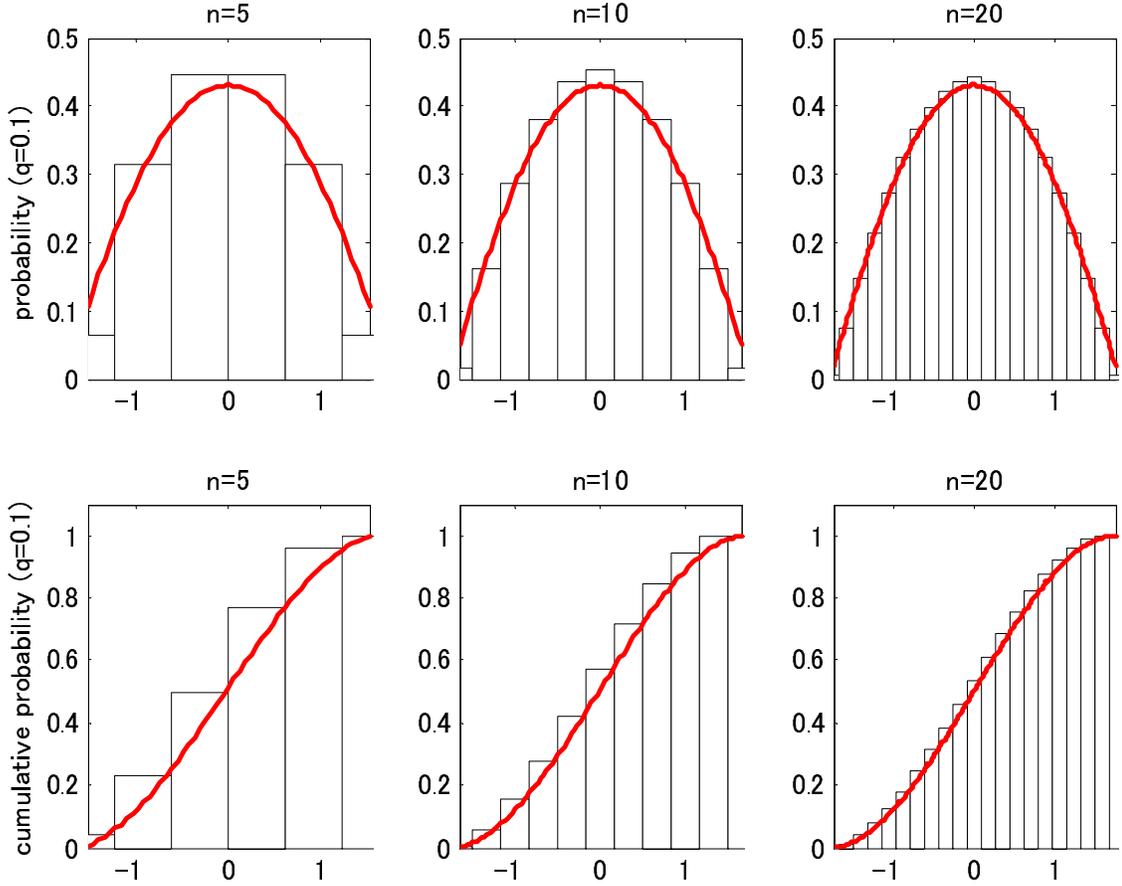}%
\caption{probability distributions (the first row) and its corresponding
cumulative probability distributions (the second row) of normalized
$q$-binomial coefficient $(q=0.1)$ and Tsallis ditribution when $n=5,10,20$}%
\label{fig: q=0.1 pdf of q-bino}%
\end{center}
\end{figure}
\begin{figure}
[ptbhptbh]
\begin{center}
\includegraphics[
height=5.2235in,
width=6.9548in
]%
{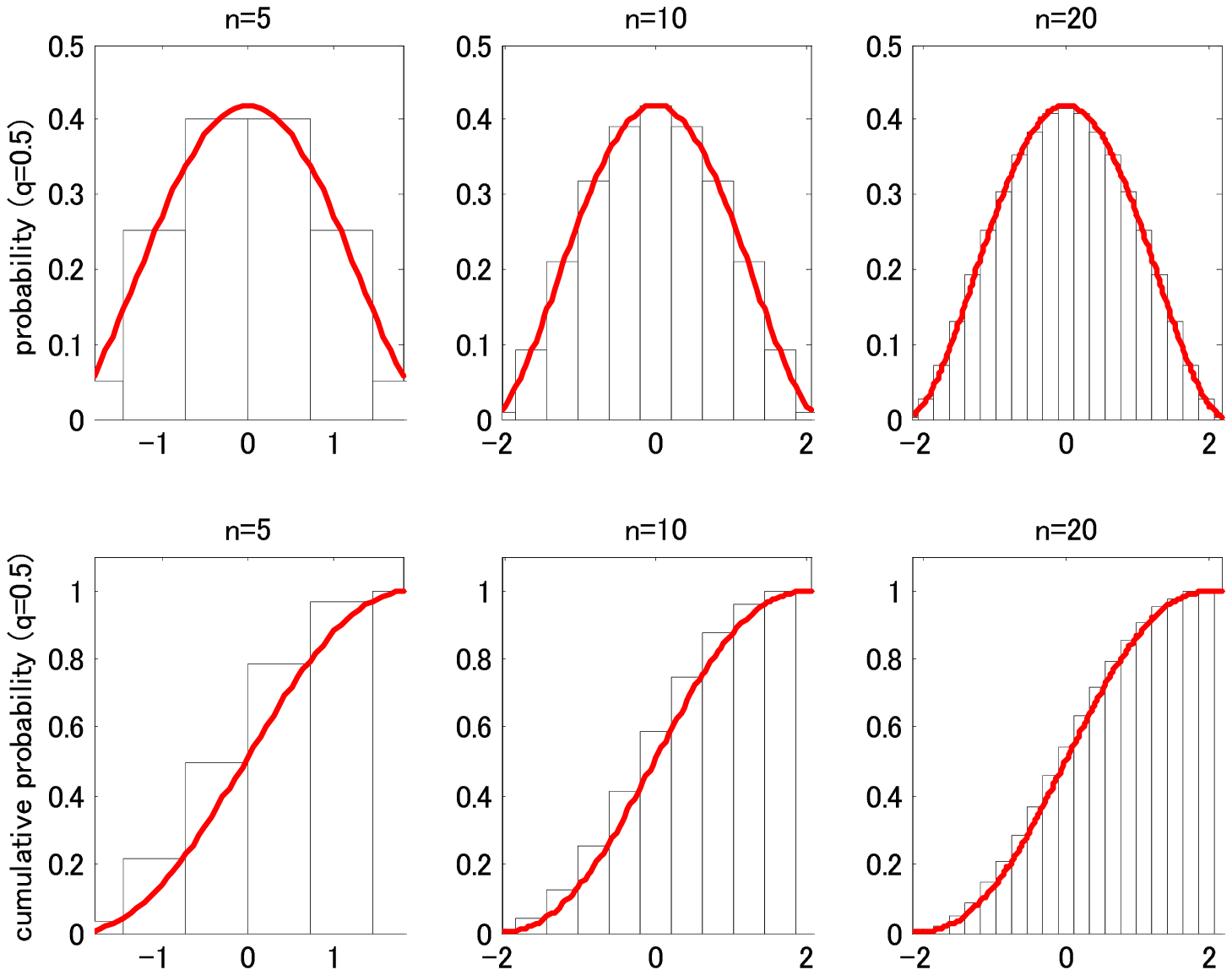}%
\caption{probability distributions (the first row) and its corresponding
cumulative probability distributions (the second row) of normalized
$q$-binomial coefficient $(q=0.5)$ and Tsallis ditribution when $n=5,10,20$}%
\label{fig: q=0.5 pdf of q-bino }%
\end{center}
\end{figure}
\begin{figure}
[ptbhptbhptbh]
\begin{center}
\includegraphics[
height=5.2235in,
width=6.9548in
]%
{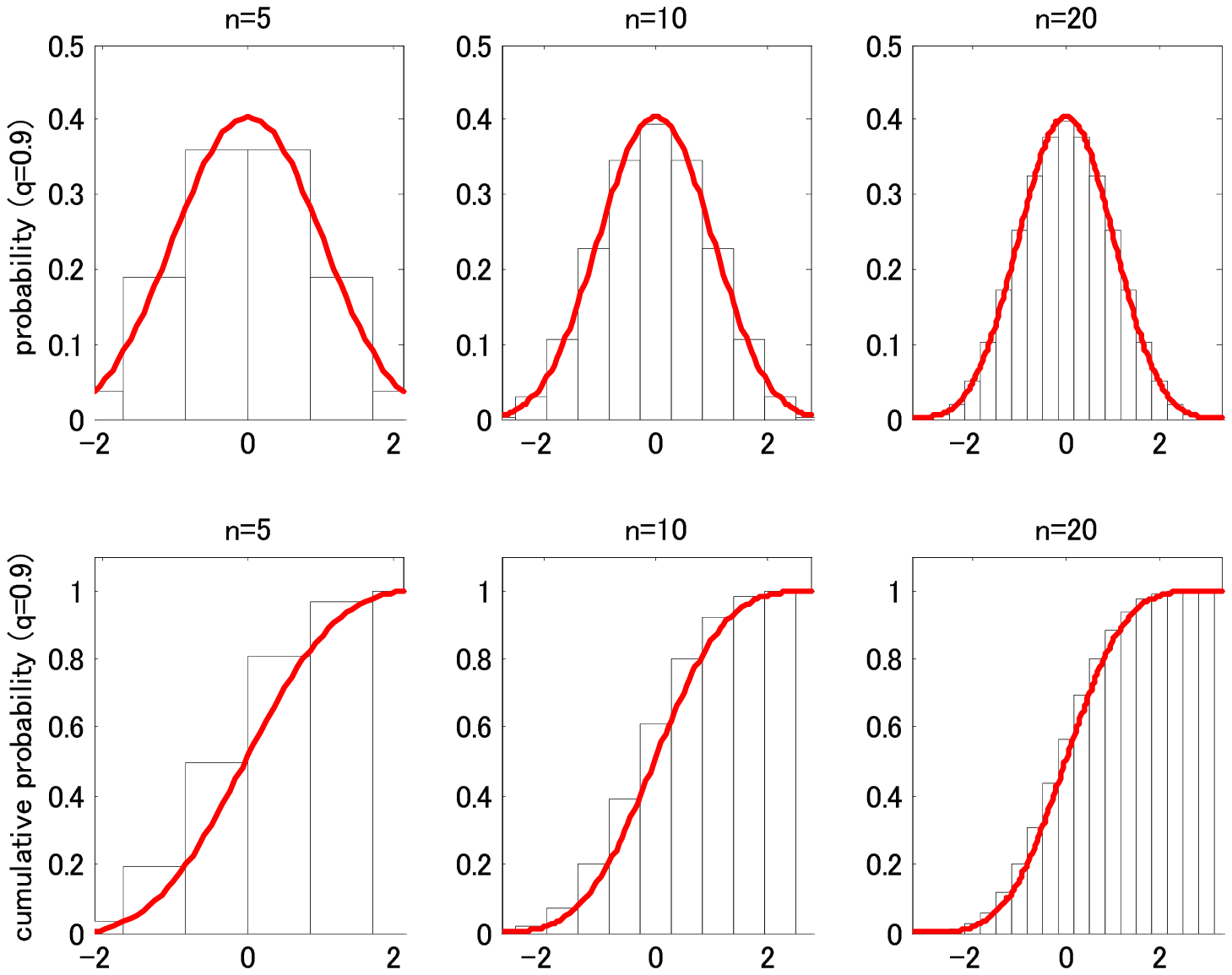}%
\caption{probability distributions (the first row) and its corresponding
cumulative probability distributions (the second row) of normalized
$q$-binomial coefficient $(q=0.9)$ and Tsallis ditribution when $n=5,10,20$}%
\label{fig: q=0.9 pdf of q-bino }%
\end{center}
\end{figure}

In order to confirm these convergences more precisely, we compute the maximal
difference $\Delta_{q,n}$ among the values of two cumulative probabilities (a
set of normalized $q$-binomial coefficients and Tsallis distribution) for each
$q=0.1,\,0.2,\,\cdots,\,0.9$ and $n.$ $\Delta_{q,n}$ is defined by%
\begin{equation}
\Delta_{q,n}:=\underset{i=0,\cdots,n}{\max}\left\vert F_{q\text{-bino}}\left(
i\right)  -F_{\text{Tsallis}}\left(  i\right)  \right\vert
\end{equation}
where $F_{q\text{-bino}}\left(  i\right)  $ and $F_{\text{Tsallis}}\left(
i\right)  $ are cumulative probability distributions of a set of normalized
$q-$binomial coefficients $p_{q\text{-bino}}\left(  k\right)  $ and its
corresponding Tsallis distribution $f_{\text{Tsallis}}\left(  x\right)  $,
respectively.%
\begin{equation}
F_{q\text{-bino}}\left(  i\right)  :=\sum_{k=0}^{i}p_{q\text{-bino}}\left(
k\right)  ,\quad F_{\text{Tsallis}}\left(  i\right)  :=\int_{-\infty}%
^{i}f_{\text{Tsallis}}\left(  x\right)  dx
\end{equation}

Fig.5 results in convergences of $\Delta_{q,n}$ to $0$ when $n\rightarrow
\infty$ for $q=0.1,\,0.2,\,\cdots,\,0.9$. This result indicates that the limit
of every convergence is a Tsallis distribution with the same $q\in\left(
0,1\right]  $ as that of a given set of normalized $q-$binomial coefficients.%
\begin{figure}
[ptbh]
\begin{center}
\includegraphics[
height=4.1987in,
width=5.2243in
]%
{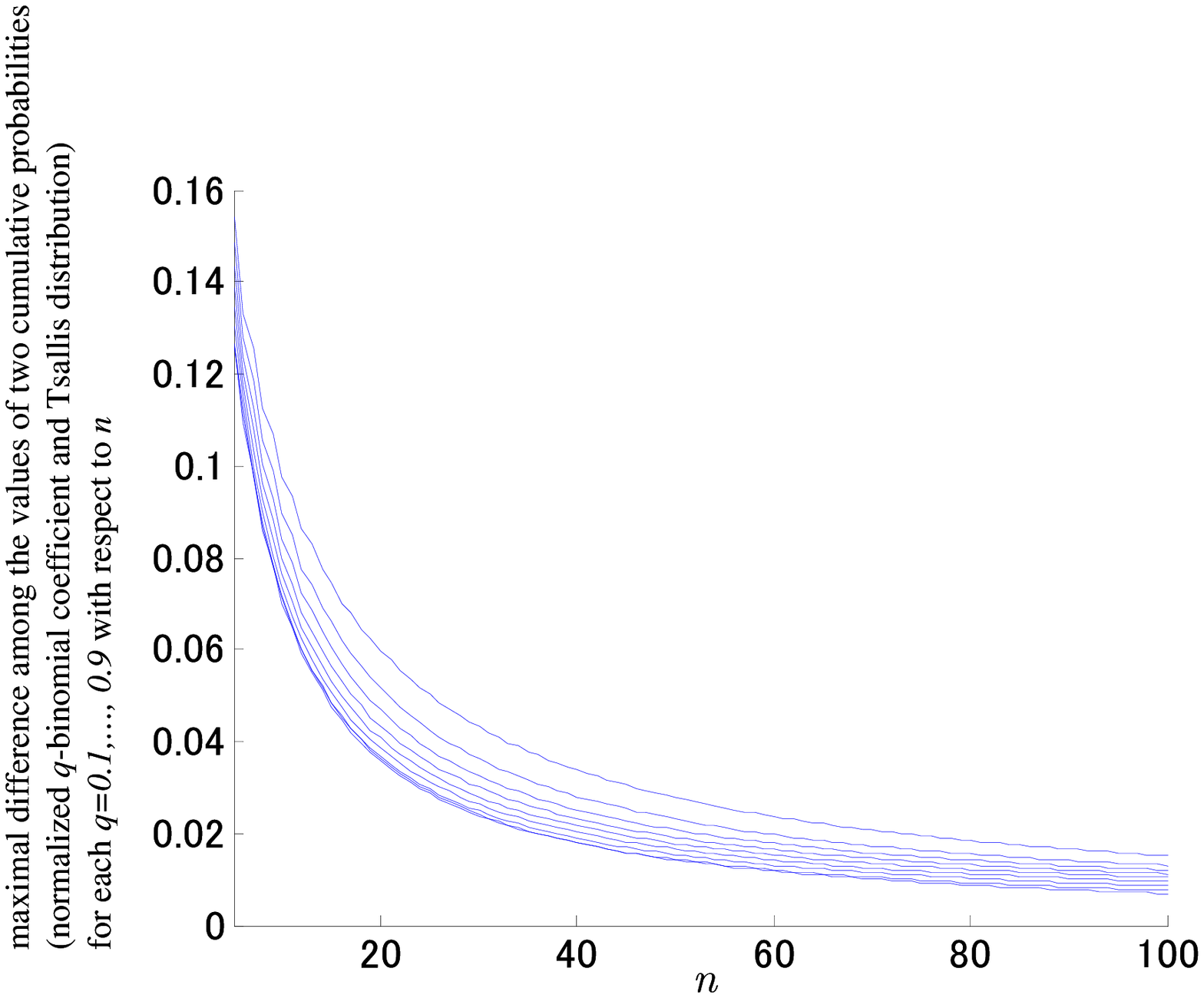}%
\caption{transition of maximal difference among the values of two cumulative
probabilities (normalized $q$-binmial coefficient and Tsallis distribution)
for each $q=0.1,0.2,\cdots,0.9$ when $n\rightarrow\infty$}%
\label{fig: max of difference}%
\end{center}
\end{figure}

The present convergences reveal a possibility of the existence of the central
limit theorem in Tsallis statistics, in which any distributions converge to a
Tsallis distribution as nonextensive generalization of a Gaussian
distribution. The central limit theorem in Tsallis statistics provides not
only a mathematical result in Tsallis statistics but also the physical reason
why there exist universally power-law behaviors in many physical systems. In
other words, the central limit theorem in Tsallis statistics mathematically
explains the reason of ubiquitous existence of power-law behaviors in nature.
\bigskip

\section{Self-similarity of the $q$-product $\otimes_{q}$ leading to
successful applications in Tsallis statistics}

In this section, we consider the reason why the $q$-product $\otimes_{q}$
provides us with successful descriptions and applications in Tsallis statistics.

In the original paper \cite{Bo03} by Borges, not only the $q$-product
$\otimes_{q}$ but also the $q$-sum $\oplus_{q}$ is introduced in the similar
way as the $q$-product $\otimes_{q}.$ The $q$-sum $\oplus_{q}$ is defined as%
\begin{equation}
x\oplus_{q}y:=x+y+\left(  1-q\right)  xy\qquad\left(  x,y\in\mathbb{R}%
,q\in\mathbb{R}^{+}\right)  .
\end{equation}
The definition of the $q$-sum $\oplus_{q}$ also originates from the following
requirements \cite{Bo03}:%
\begin{align}
\ln_{q}\left(  xy\right)   &  =\ln_{q}\left(  x\right)  \oplus_{q}\ln
_{q}\left(  y\right)  ,\\
\exp_{q}\left(  x\right)  \exp_{q}\left(  y\right)   &  =\exp_{q}\left(
x\oplus_{q}y\right)  .
\end{align}
\ The way to define the $q$-sum is very similar to that of the $q$-product,
but until now we cannot find successful applications of the $q$-sum in marked
contrast to those of the $q$-product. The reason can be considered to be
\textit{the absence of self-similarity} in the $q$-sum. On the other hand, the
operation of the $q$-product \textit{always preserves self-similarity}. Let
$f$ and $g$ be any functions which are proportional to the $q$-exponential
function (\ref{def of expq}) such as%
\begin{align}
f\left(  x\right)   &  =C\exp_{q}\left(  x\right)  =C\left[  1+\left(
1-q\right)  x\right]  ^{\frac{1}{1-q}},\\
g\left(  y\right)   &  =D\exp_{q}\left(  y\right)  =D\left[  1+\left(
1-q\right)  y\right]  ^{\frac{1}{1-q}}%
\end{align}
where $C$ and $D$ are constants. Then, we compute $f\otimes_{q}g$ and
$f\oplus_{q}g$, respectively.%
\begin{align}
f\left(  x\right)  \otimes_{q}g\left(  y\right)   &  =\left[  \left(
C^{1-q}+D^{1-q}-1\right)  +\left(  1-q\right)  \left(  C^{1-q}x+D^{1-q}%
y\right)  \right]  ^{\frac{1}{1-q}},\label{output1}\\
f\left(  x\right)  \oplus_{q}g\left(  y\right)   &  =C\left[  1+\left(
1-q\right)  x\right]  ^{\frac{1}{1-q}}+D\left[  1+\left(  1-q\right)
y\right]  ^{\frac{1}{1-q}}\nonumber\\
&  +CD\left[  1+\left(  1-q\right)  x+\left(  1-q\right)  y+\left(
1-q\right)  ^{2}xy\right]  ^{\frac{1}{1-q}}.\label{output2}%
\end{align}
Compare these outputs (\ref{output1}) and (\ref{output2}) of the two
operations $\otimes_{q}$ and $\oplus_{q}$. Obviously, the output of the
$q$-product has the same form as the $q$-exponential function
(\ref{def of expq}), but that of the $q$-sum is \textit{not} so. In other
words, the $q$-product is \textit{self-similar operation} with respect to the
$q$-exponential function, but the $q$-sum lacks such a nice property. The
presence of \textit{self-similarity} in the $q$-product results in successful
applications in Tsallis statistics. In fact, the present formulations such as
(\ref{q-kaijyo}), (\ref{expr of q-binomial}) and (\ref{expr of q-multinomial})
have the same exponential factor $\frac{1}{1-q}$. The \textit{self-similarity}
in the $q$-product makes sense in Tsallis statistics.

\section{Conclusion}

We discovered the conclusive and beautiful mathematical structure behind
Tsallis statistics. The one-to-one correspondence (\ref{important0-2}) between
the $q$-multinomial coefficient and Tsallis entropy provides us with the
significant mathematical structure such as equivalence (\ref{l=2}) in MEP,
symmetry (\ref{symmetry}), and the determination of the applicable range of
the parameter $q$ of Tsallis entropy ((\ref{range1}) and (\ref{range2})).
Moreover, the convergence of each set of $q$-binomial coefficients to each
Tsallis distribution with the same $q$ when $n$ increases represents the
existence of the central limit theorem in Tsallis statistics. These results
convince us of a possibility of the generalization of the usual probability
theory along the lines of Tsallis statistics. This means that any scientific
fields based on the usual probability theory such as statistical mechanics and
information theory can be nonextensively generalized to the nonextensive
systems for the systematical descriptions of power-law behaviors \cite{Su02a}.
These successful descriptions and applications in Tsallis statistics are
considered to originate from the presence of \textit{self-similarity} in the
$q$-product.

In all of our recently presented results such as law of error \cite{ST0312},
$q$-Stirling formula \cite{ST04a}, the representation of the $q$-exponential
function by means of the $q$-product \cite{ST04a} and the present mathematical
structure in this paper, the $q$-product is found to play crucial roles in
these successful applications. The $q$-product is uniquely determined by
Tsallis entropy. Therefore, the introduction of Tsallis entropy provides the
wealthy foundations in mathematics and physics as well-organized
generalization of the Boltzmann-Gibbs-Shannon theory. Our theoretical results
justify Tsallis statistics as a novel science for complex systems.

\begin{acknowledgement}
The author would like to thank Professor Yoshinori Uesaka and Professor Makoto
Tsukada for their valuable comments and discussions.
\end{acknowledgement}

\appendix

\section{Pascal's triangle in Tsallis statistics}

Once the $q-$binomial coefficient (\ref{def of q-binomial coefficient}) is
well-defined, we can construct Pascal's triangle in Tsallis statistics in Table.1.%

\begin{table}[tbp] \centering
\begin{tabular}
[c]{lllllllllllll}
&  &  &  &  &  & $\left[
\begin{array}
[c]{c}%
0\\
0
\end{array}
\right]  _{q}$ &  &  &  &  &  & \\
&  &  &  &  & $\left[
\begin{array}
[c]{c}%
1\\
0
\end{array}
\right]  _{q}$ &  & $\left[
\begin{array}
[c]{c}%
1\\
1
\end{array}
\right]  _{q}$ &  &  &  &  & \\
&  &  &  & $\left[
\begin{array}
[c]{c}%
2\\
0
\end{array}
\right]  _{q}$ &  & $\left[
\begin{array}
[c]{c}%
2\\
1
\end{array}
\right]  _{q}$ &  & $\left[
\begin{array}
[c]{c}%
2\\
2
\end{array}
\right]  _{q}$ &  &  &  & \\
&  &  & $\left[
\begin{array}
[c]{c}%
3\\
0
\end{array}
\right]  _{q}$ &  & $\left[
\begin{array}
[c]{c}%
3\\
1
\end{array}
\right]  _{q}$ &  & $\left[
\begin{array}
[c]{c}%
3\\
2
\end{array}
\right]  _{q}$ &  & $\left[
\begin{array}
[c]{c}%
3\\
3
\end{array}
\right]  _{q}$ &  &  & \\
&  & $\left[
\begin{array}
[c]{c}%
4\\
0
\end{array}
\right]  _{q}$ &  & $\left[
\begin{array}
[c]{c}%
4\\
1
\end{array}
\right]  _{q}$ &  & $\left[
\begin{array}
[c]{c}%
4\\
2
\end{array}
\right]  _{q}$ &  & $\left[
\begin{array}
[c]{c}%
4\\
3
\end{array}
\right]  _{q}$ &  & $\left[
\begin{array}
[c]{c}%
4\\
4
\end{array}
\right]  _{q}$ &  & \\
& $\left[
\begin{array}
[c]{c}%
5\\
0
\end{array}
\right]  _{q}$ &  & $\left[
\begin{array}
[c]{c}%
5\\
1
\end{array}
\right]  _{q}$ &  & $\left[
\begin{array}
[c]{c}%
5\\
2
\end{array}
\right]  _{q}$ &  & $\left[
\begin{array}
[c]{c}%
5\\
3
\end{array}
\right]  _{q}$ &  & $\left[
\begin{array}
[c]{c}%
5\\
4
\end{array}
\right]  _{q}$ &  & $\left[
\begin{array}
[c]{c}%
5\\
5
\end{array}
\right]  _{q}$ & \\
$\vdots$ &  & $\vdots$ &  & $\vdots$ &  & $\vdots$ &  & $\vdots$ &  & $\vdots$
&  & $\vdots$%
\end{tabular}
\caption{Pascal's triangle in Tsallis statistics\label{key}}%
\end{table}%

\bigskip The properties of the $q-$binomial coefficient are almost the same as
the usual binomial coefficient:

\begin{enumerate}
\item
\begin{equation}
\left[
\begin{array}
[c]{c}%
n\\
0
\end{array}
\right]  _{q}=1,\quad\left[
\begin{array}
[c]{c}%
n\\
1
\end{array}
\right]  _{q}=n\quad\text{for any }q\in\left[  0,1\right)  \ \text{and}%
\ n\in\mathbb{N},
\end{equation}

\item
\begin{equation}
\left[
\begin{array}
[c]{c}%
n\\
k
\end{array}
\right]  _{q}=\left[
\begin{array}
[c]{c}%
n\\
n-k
\end{array}
\right]  _{q}\quad\text{for any }q\in\left[  0,1\right]  ,n\in\mathbb{N}%
\ \text{and}\ k=0,\cdots,n,
\end{equation}

\item
\begin{equation}
k\otimes_{q}\left[
\begin{array}
[c]{c}%
n\\
k
\end{array}
\right]  _{q}=n\otimes_{q}\left[
\begin{array}
[c]{c}%
n-1\\
k-1
\end{array}
\right]  _{q}\quad\text{for any }q\in\left[  0,1\right]  ,n\in\mathbb{N}%
\ \text{and}\ k=0,\cdots,n.
\end{equation}
This is equivalent to
\begin{equation}
\left[
\begin{array}
[c]{c}%
n\\
k
\end{array}
\right]  _{q}=\left(  n\oslash_{q}k\right)  \otimes_{q}\left[
\begin{array}
[c]{c}%
n-1\\
k-1
\end{array}
\right]  _{q},
\end{equation}
which is a recursive formula of the $q-$binomial coefficient.
\end{enumerate}

\bigskip However, the important property in the usual Pascal's triangle does
not hold in general.%

\begin{equation}
\left[
\begin{array}
[c]{c}%
n\\
k
\end{array}
\right]  _{q}\neq\left[
\begin{array}
[c]{c}%
n-1\\
k-1
\end{array}
\right]  _{q}+\left[
\begin{array}
[c]{c}%
n-1\\
k
\end{array}
\right]  _{q}%
\end{equation}

Applying the explicit expression of the $q-$binomial coefficient
(\ref{expr of q-binomial}) and the above properties, Table.I is concretely
expressed as Table.II.%

\begin{table}[tbp] \centering
\begin{tabular}
[c]{lllllllllllll}
&  &  &  &  &  & $1$ &  &  &  &  &  & \\
&  &  &  &  & $1$ &  & $1$ &  &  &  &  & \\
&  &  &  & $1$ &  & $2$ &  & $1$ &  &  &  & \\
&  &  & $1$ &  & $3$ &  & $3$ &  & $1$ &  &  & \\
&  & $1$ &  & $4$ &  & $_{4}C_{2}^{\left(  q\right)  }$ &  & $4$ &  & $1$ &  &
\\
& $1$ &  & $5$ &  & $_{5}C_{2}^{\left(  q\right)  }$ &  & $_{5}C_{3}^{\left(
q\right)  }$ &  & $5$ &  & $1$ & \\
$\vdots$ &  & $\vdots$ &  & $\vdots$ &  & $\vdots$ &  & $\vdots$ &  & $\vdots$
&  & $\vdots$%
\end{tabular}
\caption{Pascal's triangle in Tsallis statistics\label{key}}%
\end{table}%
%

\begin{align}
_{4}C_{2}^{\left(  q\right)  } &  :=\left[
\begin{array}
[c]{c}%
4\\
2
\end{array}
\right]  _{q}=\left[  -2^{1-q}+3^{1-q}+4^{1-q}\right]  ^{\frac{1}{1-q}},\\
_{5}C_{2}^{\left(  q\right)  } &  :=\left[
\begin{array}
[c]{c}%
5\\
2
\end{array}
\right]  _{q}=\left[  -2^{1-q}+4^{1-q}+5^{1-q}\right]  ^{\frac{1}{1-q}},\\
_{5}C_{3}^{\left(  q\right)  } &  :=\left[
\begin{array}
[c]{c}%
5\\
3
\end{array}
\right]  _{q}=\left[  -2^{1-q}+4^{1-q}+5^{1-q}\right]  ^{\frac{1}{1-q}}.
\end{align}
Clearly, the effect of $q$ in Tsallis statistics becomes bigger in the lower
rows of Pascal's triangle. This is in good agreement with the physical models
in \cite{CLDO01}.

Note that from the theoretical results in section III the probability
distribution by normalizing the $q$-binomial coefficients $\left[
\begin{array}
[c]{c}%
n\\
k
\end{array}
\right]  _{q}$ on the $(n+1)$-th row from the top of the present Pascal's
triangle (Table I) converges to a Tsallis distribution with the same $q$ when
$n$ increases.

Using the present theoretically derived Pascal's triangle in Tsallis
statistics, the multifractals in Tsallis statistics can be investigated in
more detail, which does not depend on any physical model.

\end{document}